\documentclass[12pt]{article}
\usepackage{graphicx}
\usepackage{subfigure}
\usepackage{amsmath}
\usepackage{amssymb}

\begin{document}

\begin{center}
\begin{large}
\title\\{ \textbf{A Higher Dimensional Potential Model Approach in the Study of Isgur-Wise Function for Heavy-Light Mesons.}}\\\
\end{large}

\author\

$Sabyasachi\;\; Roy^{\emph{1}}\footnotemark \;\;and \;\;D.\;K.\;Choudhury^{\emph{1,2,3}}$  \\
\footnotetext{Corresponding author (On leave from Karimganj College,Assam, India). e-mail :  \emph{sroy.phys@gmail.com}}
\textbf{1}. Department of Physics, Gauhati University, Guwahati-781014, India.\\
\textbf{2}. Physics Academy of The North East, Guwahati-781014, India. \\
\textbf{3}. Centre for Theoretical Studies, Pandu College, Guwahati-781012, India.\\

\begin{abstract}
\begin{center}
Nambu-Goto action of bosonic string predicts the quark-antiquark potential to be $ V ( r ) = \frac{\gamma}{r} + \sigma r +\mu_0 $.  The coefficient$ \gamma = - \frac{\pi(d-2)}{24} $  is the universal L\"{u}scher coefficient of the L\"{u}scher term  $ \frac{\gamma}{r}$ , which depends upon the space-time dimension `d'. We take linear term in potential as parent and L\"{u}scher term  as perturbation for the generation of wave function for meson in d space-time dimension.The wave function comes out in terms of Airy's infinite polynomial series. With this wave function in higher dimension, we then study the Isgur-Wise function for heavy-light mesons and its derivatives. \\\

\emph{Key words : Nambu-Goto potential, L\"{u}scher Term, $D$ dimensional Schr\"{o}dinger equation. \\\
}
\end{center}

\end{abstract}
\end{center}

\section{Introduction:}\rm

It is well known that in the non-perturbative regime of QCD, i.e, in the low-energy regime, phenomenological models are very useful for the study of hadrons [1]. Since long, potential models for mesons, with one heavy and one light quarks, are under theoretical investigation for the study of  properties of hadrons like their mass, form factors, decay widths etc. \\
It is worth mentioning here that, for getting the mesonic wave-function in such potential model approach, the choice of potential is of utmost importance. Several potentials are there for modeling quark-antiquark bound states like Martin Potential[2], Cornell Potential[3], Richardson potential[4],Logarithmic potential[5]. Out of these, the Cornell potential, with \emph{linear plus Coulombic} form, has been very popular and useful for such phenomenological study in QCD. The wave function for heavy-light mesons have been calculated earlier with such a potential by applying quantum mechanical perturbation technique [6]. This has been deduced both with Coulombic term in potential as parent[7-9] and also with linear confinement term as parent[10-11]. In the present work, we report wave function for meson developed within some higher dimensional string inspired potential model approach. \\
 In classical string model for hadrons[12-14], proposed by Yoichiro Nambu and Tetsuo Goto, two-quark potential can be expressed as [15] $V ( r ) =\frac{\gamma}{r}+ \sigma r +\mu_0$. Here,the coefficient$ \gamma = - \frac{\pi(d-2)}{24} $  is the universal L\"{u}scher coefficient of the L\"{u}scher term  $ \frac{\gamma}{r}$ [16], which depends upon the space-time dimension d. $\sigma$ is the string tension which characterizes the strength of the confining force between static charges. Its value is $ 0.178\; GeV^2$ [17,18], $\mu_0$ is a regularisation dependent constant. This Nambu-Goto potential in higher dimension is analogous to the linear plus Coulombic type Cornell potential in $3+1$ dimensional QCD. \\
Earlier, we have developed similar formalism considering L\"{u}scher term as parent in perturbation technique [53]. However, physical basis of string potential model supports the fact that the linear confinement term as the main contributing part in inter-quark potential and L\"{u}scher's term arises in this potential as a first order correction to the linear confinement term  $\sigma r$ [13]. In our present approach of developing wave function for meson using perturbation technique, we consider the leading linear term as parent and L\"{u}scher term as perturbation and use D-dimensional Schrodinger equation for solving out the wave function of mesons. Several methods for solution of higher dimensional Schrodinger equation with different potentials are there in literature[19-29]. This we take as a generalisation of our previous works in three-dimensional QCD [10-11] to higher dimension. \\
However, due to the lack of exact solution for $D$ dimensional Schr\"{o}dinger equation with linear plus Coulombic type potential, we opt for employing perturbation method. With linear term in potential as parent, the wave function comes out in terms of Airy's function [30], which is an infinite polynomial series. Earlier, some approximated wave function for meson has been deduced within QCD framework, with linear term in Cornell potential as parent and coulombic term as perturbation, considering Airy's polynomial function up to $O(r^3)$ [10], which is recently further improved upon by the authors considering complete Airy series [52]. \\
In this paper, we have reported the wave function for meson in higher space-time dimension considering complete Airy's infinite series, following Dalgarno's method of perturbation[6,31-32]. Based on this wave function, we then have studied Isgur-Wise function (IWF) for heavy-light mesons [33-34] and its derivatives in higher space-time dimension. IWF is a universal function representing all the form factors of heavy-light mesons in the infinite mass limit in semileptonic transitions [35] of these mesons. It is a fundamental quantity in QCD, which can be determined non-perturbatively. \\
In such a study with wave function containing Airy's infinite series,the infinite upper limit of integration in the calculations of normalisation constant and derivatives of IWF, gives rise to divergence. In our earlier works [11] we have successfully introduced some reasonable cut-off to upper infinite limit of integration in the study of IWF and its derivatives. Here we fix the cut-off to infinite upper limit by applying the convergence condition of wave function in perturbation theory. We also study the dimensional dependence of these parameters and compare our present results with our earlier ones and also with the corresponding theoretical  and experimental expectations for $D=3$ [36-43,53].\\
With this introduction, detailed formalism is reported in section 2, calculations and results in section 3. Final conclusion and remarks are reported in section 4.

\section{Formalism:}

\subsection{Potential Model:}
As expressed in equation (1), the quark-antiquark potential for bosonic strings as enunciated by Nambu and Goto  has the standard form:
\begin{equation}
V(r)=\frac{\gamma}{r}+\sigma r +\mu_{0}
\end{equation}
with $\gamma$ as the L\"{u}scher term given by :
\begin{equation}
\gamma=- \frac{\pi(d-2)}{24}
\end{equation}
For the sake of simplification in formalism and comparison with Cornell potential [3], we take Nambu-Goto potential of equation (1) as:
\begin{equation}
V(r)=-\frac{\gamma}{r}+\sigma r +\mu_{0} \;\;now \;\; with\;\;\; \gamma = \frac{\pi(d-2)}{24}.
\end{equation}
Here,d is the space-time dimension with $d=D+1$, D is the spatial dimension. \\
We take $ \sigma r $ as parent and $ -\frac{\gamma}{r}$ as perturbation, with regularisation term $\mu_{0}=0$. Our unperturbed Hamiltonian is[19]:
\begin{equation}
H_0=-\frac{\nabla_D^{2}}{2\mu}+ \sigma r
\end{equation}
The Schrodinger equation in D-dimension is [23]:
\begin{equation}
[-\frac{\hbar^{2}}{2\mu}\nabla_D^{2} +V_0(r)]\Psi(r,\Omega_D)=E\Psi(r,\Omega_D)
\end{equation}
with [55,56],
\begin{equation}
 \nabla_D^{2}=\frac{1}{r^{D-1}}\frac{d}{dr}(r^{D-1}\frac{d}{dr})-\frac{\Lambda_D^{2}(\Omega_D)}{r^2}
 =\frac{d^{2}}{dr^{2}}+\frac{D-1}{r}\frac{d}{dr}-\frac{\Lambda_D^{2}(\Omega_D)}{r^{2}}
\end{equation}
and
\begin{equation}
\Psi(r,\Omega_D)=R(r)Y(\Omega_D)
\end{equation}
Here, $ \frac{\Lambda_D^{2}(\Omega_D)}{r^{2}}$ is a generalisation of the centrifugal barrier [44] in D dimension. \\
The eigen values of $ \Lambda_D^{2}(\Omega_D) $ age given by :
\begin{equation}
\Lambda_D^{2}(\Omega_D)Y(\Omega_D)=l(l+D-2)Y(\Omega_D)
\end{equation}
Here $Y(\Omega_D)$ and $R(r)$ are the spherical harmonics and spherical coordinates; $l$ is the angular momentum quantum number and E is the energy Eigen value. \\
This gives the equation (5) in term of radial part as :
\begin{equation}
[\frac{d^{2}}{dr^{2}}+\frac{D-1}{r}\frac{d}{dr}-\frac{l(l+D-2)}{r^{2}}+\frac{2\mu}{\hbar^{2}}(E-V_0)]R(r)=0
\end{equation}
Taking $\hbar=1 $ and for $ l=0 $ state, we have:
\begin{center}
\begin{eqnarray}
[\frac{d^{2}}{dr^{2}}+\frac{D-1}{r}\frac{d}{dr}+2\mu(E-V_0)]R(r)=0 \\\
R^{\prime\prime}(r)+\frac{D-1}{r}R^{\prime}(r)+2\mu(E-\sigma r)R(r)=0
\end{eqnarray}
\end{center}

We take:
\begin{equation}
R(r)=r^{\frac{(1-D)}{2}}U(r)
\end{equation}
Then our equation (11) transforms into :
\begin{eqnarray}
U^{\prime\prime}(r)-\frac{(D-1)(D-3)}{4r^{2}}U(r)+2\mu(E-\sigma r)U(r)=0\\
or\; U^{\prime\prime}(r)-\frac{\Lambda (\Lambda+1)}{r^{2}}U(r)+2\mu(E-\sigma r)U(r)=0\;\;with\;\; \Lambda = \frac{D-3}{2}
\end{eqnarray}

\begin{flushleft}
\textbf{Wave function with only linear term in potential :} \\
\end{flushleft}

We take :
\begin{equation}
\varrho(r)= \varrho_1 r - \varrho_0
\end{equation}
So, the equation (14) reduces to :
\begin{equation}
[\frac{d^{2}}{d\varrho^{2}}- \frac{\Lambda(\Lambda+1)}{(\varrho+\varrho_0)^{2}}+\frac{2\mu E}{\varrho_1 ^{2}}-\frac{2\mu \sigma \varrho}{\varrho_1 ^{2}}-\frac{2\mu\sigma \varrho_0}{\varrho_1 ^{3}}]U(\varrho)=0
\end{equation}
\textbf{\emph{Case-I:}} \\
For the case $ \varrho\rightarrow \infty $ , $ \frac{1}{(\varrho + \varrho_0 )^{2}} $ term vanishes, the above equation (16) becomes :
\begin{equation}
[\frac{d^{2}}{d\varrho^{2}}+\frac{2\mu E}{\varrho_1 ^{2}}-\frac{2\mu \sigma \varrho}{\varrho_1 ^{2}}-\frac{2\mu\sigma \varrho_0}{\varrho_1^{3}}]U(\varrho)=0
\end{equation}
Following our previous work [11], we take $ \varrho_1 = (2\mu\sigma )^{1/3}$ and $\varrho_0 =(\frac{2\mu}{\sigma^2})^{1/3} E $,
so that our construction of $ \varrho(r) $ becomes:
\begin{equation}
\varrho(r)= \varrho_1 r - \varrho_0 = (2\mu\sigma )^{1/3} r - (\frac{2\mu}{\sigma^2})^{1/3} E
\end{equation}
Then equation (17) transforms into:
\begin{equation}
\frac{d^{2}U(\varrho)}{d\varrho^{2}}-\varrho U(\varrho)= 0
\end{equation}
Here, we mention that, $ \varrho_0$ are the zeroes of the Airy's function such that $Ai[\varrho_0]=0$. The solution of the equation (19) comes out in terms of Airy's function $ Ai[\varrho] $, as[45]:
\begin{equation}
U(\varrho)\sim Ai[\varrho] = NAi[(2\mu\sigma )^{1/3} r - (\frac{2\mu}{\sigma^2})^{1/3} E]
\end{equation}
This is the asymptotic solution of equation (17).\\
\textbf{\emph{Case-II:}}\\
If we now take $\varrho\rightarrow 0 $, then the centrifugal term of $ \frac{1}{(\varrho + \varrho_0 )^{2}} $  will be dominant so that equation (16) can be reduced as :
\begin{equation}
U^{\prime\prime}(r)-\frac{\Lambda (\Lambda+1)}{(\varrho+\varrho_0)^{2}}U(\varrho)=0
\end{equation}
The non-singular solution of this equation is :
\begin{equation}
U(\varrho)\sim (\varrho +\varrho_0 )^{\sqrt{\Lambda(\Lambda+1)}}
\end{equation}
We construct the approximate analytic solution of equation (16) as the product of two solution of extreme cases [46] as:
\begin{equation}
U(\varrho)\sim (\varrho +\varrho_0)^m Ai[\varrho]
\end{equation}
The unperturbed ground state meson wave function with linear term in potential, has thus the form :
\begin{equation}
\Psi^{0}(r,D)=N r^{\frac{(1-D)}{2}} (\varrho_1 r)^{m}Ai[\varrho_1 r-\varrho_0],\;\;  m={\sqrt{\Lambda(\Lambda+1)}}
\end{equation}
Here, N is the normalisation constant for unperturbed wave function. \\
\begin{flushleft}
\textbf{Wave function with L\"{u}scher term as perturbation:} \\
\end{flushleft}
Given this unperturbed wave function in eqn (24), we now proceed to construct the perturbed (and hence total) wave function for meson taking L\"{u}scher term in potential as perturbation. \\
Our D-dimensional Schrodinger equation is:
\begin{equation}
[-\frac{1}{2\mu}(\frac{d^{2}}{dr^{2}}+\frac{D-1}{r}\frac{d}{dr})+(\sigma r -E )]\Psi^{\prime}(r)=(\frac{\gamma}{r}+W^{\prime})\Psi^{0}(r,D)
\end{equation}
Here $W^{\prime}$ is the first order perturbed energy eigenvalue and in $D$ spatial dimension [54], it is having the expression :
\begin{equation}
W^{\prime}=-\int_0 ^\infty D C_D r^{D-1} |\Psi(r)|^2 \frac{\gamma}{r}dr
\end{equation}
where $C_D=\frac{\pi^{D/2}}{\Gamma(\frac{D}{2}+1)}$. \\
In terms of radial wave function $R_1(r,D)$ the equation (25) can now be expressed as:
\begin{equation}
[-\frac{1}{2\mu}(\frac{d^{2}}{dr^{2}}+\frac{D-1}{r}\frac{d}{dr})+(\sigma r -E )]R_1(r,D)=(\frac{\gamma}{r}+W^{\prime})r^{\frac{(1-D)}{2}} (\varrho_1 r)^{m}Ai[\varrho_1 r-\varrho_0]
\end{equation}
We assume,
\begin{equation}
R_1(r,D)= r^{\frac{(1-D)}{2}}F(r,D) (\varrho_1 r)^{m}Ai[\varrho_1 r-\varrho_0]
\end{equation}
Employing Dalgarno's method of perturbation the unperturbed wave function comes out as (Appendix-A):

\begin{equation}
\Psi^{\prime}(r,D)=N_1r^{\frac{(1-D)}{2}}(\varrho_1 r)^{m}[A_1(r,D)r+A_2(r,D)r^{2}+A_3(r,D)r^{3}+.........]Ai[\varrho_1 r-\varrho_0]
\end{equation}
$A_1(r)$ , $A_2(r)$, $A_3(r)$ etc are having explicit form as given by equation (A.24-A.28) of Appendix-A.
Our the total wave function comes out to be:
\begin{equation}
\Psi^{tot}(r,D)=\Psi^{0}(r)+\Psi^{\prime}(r)= N_1r^{\frac{(1-D)}{2}}[1+A_1(r,D)r+A_2(r,D)r^{2}+A_3(r,D)r^{3}+.........](\varrho_1 r)^{m}Ai[\varrho_1 r-\varrho_0]
\end{equation}
Here, $N_1$ is the normalisation constant for the total wave function. \\

\subsection{Isgur-Wise Function and its derivatives:}
For meson containing one heavy quark (c,b,t), the mass of heavy quark is much greater than the QCD scale parameter $\Lambda_{QCD}$; the four-velocity of heavy quark is almost same as the four velocity of meson  and in the meson rest frame, the heavy quark appears as a static colour source[47]. This brings spin flavour symmetry and reduction in the number of form factors[48]. All the six form factors in semi-leptonic decay are now expressible in terms of a single universal function depending only on the velocity transfer. This is known as the Isgur-Wise Function. It measures the overlap of of the wave function of light degrees of freedom in the initial and final mesons. This IWF $\xi(v,v^{\prime})$ is a function of four velocities $v$ and $v^{\prime}$ of the heavy particle before and after decay.
As $E_{recoil}=\mu(v.v^{\prime}-1)$ is the recoil energy of the final state meson $P^{\prime}$ (of mass $\mu$) in the rest frame of initial meson $P$, the point $y=v.v^{\prime}=1$ is referred to as the zero recoil limit. The IWF is thus normalised at zero recoil point, which is a consequence of current conservation [57].If $y=v.v^{\prime}$ (known as Lorentz boost), then, for zero recoil ($y=1$), $\xi(y)=1$. Knowledge of IWF and its derivatives is essential in all calculations that lead to the estimation of branching ratios of semi-leptonic decays and of the elements of CKM matrix. \\
The modeling of IWF is guided by the condition of its extrapolation to zero recoil. Actually there is no unique formulation of IWF and different ansatze produce different expressions of IWF [49]. These expressions differ at large values of $y$, but agreeing within the range $y\approx 1$. Generally, the available experimental data fit to linear form of IWF [$\xi(y)=1-\rho^2 (y-1)$], cosisting of only first order derivative term. But, as $y$ increases, higher order terms make difference, as far as the shape of the IWF is concerned. For larger values of slope parameter ($\rho^2 > 1$), it is important to include higher terms, in the expansion of IWF about $y=1$, in the analysis of experimental data. Also, retaining only the first term in expansion leads to underestimate of slope parameter, making it essential to include the quadratic term (curvature) to get reliable results. \\
Also, it is very important to have theoretical insight on the expansion parameters for extrapolation to $y=1$. As, present experimental and theoretical analysis by and large concentrate on results up to slope and curvature, we confine our formulation of IWF up to quadratic term in the expansion. This is pretty well acceptable for $y<1.2$ [58]. Under this consideration, for small non-zero recoil, IWF can be expressed as [49]:
\begin{equation}
\xi(y)=1-\rho^2 (y-1) +C(y-1)^2 + \cdots\cdots
\end{equation}
Here $\rho^2$ is the slope (charge radii) and $C$ is the curvature (convexity parameter) of IWF, which are measured at zero recoil point as:
\begin{equation}
\rho^2 = -\frac{\delta\xi (y)}{\delta y}|_{y=1} \;\;,\;\; C=\frac{\delta^{2}\xi (y)}{\delta y^2}|_{y=1}
\end{equation}
The calculation of $ \rho^{2}$ and $C $ provides a measure of the validity of HQET  in the infinite quark mass limit. There have been several attempts to calculate $ \rho^{2}$ and $C$ from theory and models[36-43]. The corresponding results are shown in Table-9. The curvature parameter $C$ gives the measure and direction of convexity of the the shape of IWF. A negative curvature would result in concave IWF in the plot. However, on general ground, the IWF is expected to have positive curvature for all $y>1$. A quantitative reason in calculating curvature along with slope parameter is that beyond the first derivative, higher derivatives play a non-negligible role at $y>1$, as discussed earlier. Also, as data are now coming out to be more and more precise, it is not only of an academic interest to analyze higher derivatives of IWF. \\
In non-relativistic quark model (NRQM), IWF describes the overlap between wave functions of light degrees of freedom. The calculation of this IWF  is non-perturbative in principle [50]and this function depends upon the meson wave function and some kinematic factor as given below, in $D$ spatial dimension:
\begin{equation}
\xi(y)=\int_0 ^\infty D C_D r^{D-1} |\Psi(r)|^2\cos(pr)dr
\end{equation}
Here, $p^2$ is the square of virtual momentum transfer. Since the hadronic matrix element describing the weak decay process is invariant under complex conjugation along with an interchange of $v$ and $v^{\prime}$, we get the squared invariant velocity transfer as $(v - v^{\prime})^2 = 2(y-1)$[57]. From this we get $ p^2=2\mu^2 (y-1)$. From current conservation, it is now obvious that IWF is normalised at $p^2 =0$. \\
As $\cos(pr)=1-\frac{p^2 r^2}{2}+\frac{p^4 r^4}{4}$ +$\cdot\cdot\cdot\cdot\cdot\cdot$, considering $\cos(pr)$ up to  $O(r^4)$ we get,
\begin{equation}
\xi(y)= D C_D\int_0 ^\infty  r^{d-1} |\Psi(r)|^2dr - [D C_D\mu^2\int_0^\infty r^{D+1}|\Psi(r)|^2dr](y-1)+[\frac{1}{6}D C_D\mu^4\int_0^\infty r^{D+3}|\Psi(r)|^2dr](y-1)^2
\end{equation}
Comparing (31) and (34) give us :
\begin{eqnarray}
\rho^2 = D C_D\mu^2\int_0^\infty r^{D+1}|\Psi(r)|^2dr \\
C= \frac{1}{6}D C_D\mu^4\int_0^\infty r^{D+3}|\Psi(r)|^2dr \\
D C_D\int_0 ^\infty r^{D-1} |\Psi(r)|^2dr =1
\end{eqnarray}
Equation (37) gives the normalization constants $ N \;\;$ and $\;\;  N_1$ for $\Psi^0 (r,D)$ and $\Psi^{total} (r,D)$.

\section{Calculations and Results:}\rm
With the wave function constructed, we now proceed to study IWF and its derivatives like slope and curvature parameters. From expression of total wave function in equation (30), we find that it contains two infinite series - one of power series in r with coefficients $A_1(r,D), A_2(r,D),A_3(r,D)$ etc and another of infinite Airy series $Ai[\varrho]$. As the infinite limit of integration in the calculations of IWF and its derivatives, involving Airy's infinite series, gives rise to divergence, we opt for some reasonable cut-off to this infinite upper limit of integration. In principle, this cut-off $r_0$ should be greater than the size of hadrons ,i.e, $r_0$ should be greater than $\frac{1}{m_h}$, $m_h$ being the mass of hadron. The convergence condition of the total wave function obtained through perturbation technique implies that $\Psi^{\prime}(r) < \Psi(r) $. From this we obtain:
\begin{equation}
\mid A_1(r,D) r + A_2(r,D) r^2 +A_3(r,D) r^3 +\cdots\cdots \mid < 1
\end{equation}
This condition gives us the limiting values of cut-off $r_0$ for different $D$ values. Considering two terms in equation (38), we obtain the equation relating cut-off $r_0$ and dimension $D$ as:
\begin{equation}
A_1(r_0,D) r_0 + A_2(r_0,D) r_0^2 =1
\end{equation}
Using expressions for $A_1(r,D)$,$A_2(r,D)$ and on simplification, equation (39) transforms into polynomial series in $r_0$:
\begin{equation}
K_{11}(D) r_0^3 +K_{12}(D)r_0^2 +K_{13}(D) r_0 +K_{14}(D) =0
\end{equation}
Here, coefficients $K_{11}(D),K_{12}(D)$ etc are independent of $r_0$ , having the following explicit D-dependence in terms of the function $f(D)=\frac{1-D+2m}{2}$.
\begin{center}
\begin{flushleft}
\begin{eqnarray}
K_{11}(D)=2\mu W^{\prime}[f(D)(f(D)-1)-(D-1)f(D)+2f(D)+(D-1)\varrho_1]\\
K_{12}(D)=-4\mu \gamma [2 f(D) +(D-1)+ 2 \varrho_1] + 2\mu W^{\prime} [2 \varrho_1 k] \nonumber \\
 - 2\mu\gamma [f(D)[f(D)-1]-(D-1)f(D)+2f(D)+(D-1)\varrho_1] +2\mu W^{\prime} [(D-1)\varrho_1 k ] \\
K_{13}(D)= 2\mu W^{\prime} [2 f(D) +(D-1)+ 2 \varrho_1]- 4\mu\gamma [2 \varrho_1 k] \nonumber \\
 -2\mu\gamma [(D-1)\varrho_1 k ]+2\mu W^{\prime} [\varrho_1^2 k^2]+1 \\
K_{14}(D)= -2\mu\gamma [\varrho_1^2 k^2]
\end{eqnarray}
\end{flushleft}
\end{center}
Equation (40) is a cubic equation having only one real root which is given by :
\begin{flushleft}
\begin{eqnarray}
r_0=-\frac{K_{12}}{3K_{11}}-\frac{2^{1/3}(-K_{12}^2+3K_{11} K_{13})}{3K_{11} g^{1/3}}+\frac{1}{3.2^{1/3}K_{11}}g^{1/3} \\
with \;\; g=9K_{11} K_{12} K_{13}-2K_{12}^3-27 K_{11}^2 K_{14}+  \nonumber \\
\sqrt{4(3K_{11} K_{13}-K_{12}^2)^3 +(9 K_{11} K_{12} K_{13} -2K_{12}^3-27 K_{11}^2 K_{14})^2} \nonumber
\end{eqnarray}
\end{flushleft}
Following equation (45)and (41-44), the numerical values of cut-off $r_0$  for D and B mesons for different dimensions are shown in Table-1.

\begin{table}[!htbp]
\begin{center}
\caption{Values of $r_0$ for different $D$}
\begin{tabular}{|cc||cc|}
  \hline
    D meson & &  B meson & \\
   \hline
  D & $r_0$  & D & $r_0$  \\
   \hline \hline
  3  & 3.6354    &   3  & 3.4719   \\
  4  & 3.9482    &   4  & 3.7678   \\
  5  & 4.1361    &   5  & 3.9437     \\
  9  & 4.6362    &   9  & 4.4135     \\
  15 & 5.1158    &   15 & 4.8657  \\
  20 & 5.4104    &   20 & 5.1436   \\
  25 & 5.6515    &   25 & 5.3712   \\
  \hline
\end{tabular}
\end{center}
\end{table}

Also,consideration of such cut-off to upper limit of integrations will not sacrifice the nature and value of IWF and its derivatives, because, Airy's function falls very sharply (almost exponentially) and almost dies out with increasing r-value [45,51]. In fact, the Airy's function value becomes negligibly small for $r > 4$ ($AiryAi[4]=0.000952 $). Further, we restrict our calculation up to Airy order $r^{10}$ and term $A_5(r,D)$ in the second infinite series. The values of unperturbed energy E and of perturbed energy $W^{\prime}$ for different $D$ value are shown in Table-2. The table reflects that perturbed energy $W^{\prime}$ is less than the unperturbed energy $E$, as it should be.\\
\begin{table}[!htbp]
\begin{center}
\caption{Values of $E$ and $W^{\prime}$ in GeV for different $D$}
\begin{tabular}{|ccc||ccc|}
  \hline
   & D meson & & & B meson & \\
   \hline
  D & $E$ & $W^{\prime}$ & D & $E$ & $W^{\prime}$ \\
   \hline \hline
  3  & 0.345938 & 0.137536   &   3  & 0.352764 & 0.139463  \\
  4  & 0.637435 & 0.141397   &   4  & 0.708288 & 0.143216  \\
  5  & 0.832819 & 0.148987   &   5  & 1.0102   & 0.149692  \\
  9  & 1.54202  &  0.163230  &   9  & 1.66759  & 0.16872   \\
  15 & 1.73322  & 0.195488   &   15 & 1.74195  & 0.20805   \\
  20 & 1.7477   & 0.240184   &   20 & 1.74838  & 0.24205  \\
  25 & 1.75193  & 0.306038   &   25 & 1.76108  & 0.306265  \\
  \hline
\end{tabular}
\end{center}
\end{table}

It is to be mentioned here that, $A_1(r,D), A_2(r,D),A_3(r,D)$ etc appearing in the second infinite series of the wave function are functions of $C_1(r)$ and $C_2(r)$. Now, for the simplification of calculation, we consider lowest Airy order in computing $C_1(r)$ and $C_2(r)$[Appendix-B].\\
With these more explicit forms of functions $A_i(r,D)$ in meson wave function as in equation (30), we now explore IWF and its derivatives for Airy's polynomial order $r^{10}$, considering specific cut-off values for specific dimension value $D$. Results for B and D mesons are shown in Table-3. \\
\begin{table}[!htbp]
\begin{center}
\caption{$\rho^{2}$ and $C $ for different $D$ values.}\label{centre}
\begin{tabular}{|c|cc|cc|}
  \hline
   &  D meson & & B meson & \\
     \hline \hline
       $D$ &  $\rho^2$ & C   & $\rho^2$ & C   \\
   \hline
   3  & 0.2158 & 0.0174 & 0.2608 & 0.0254    \\    \hline
   4  & 0.3874 & 0.0457 & 0.4039 & 0.0483    \\    \hline
   5  & 0.4366 & 0.0613 & 0.4813 & 0.0679    \\    \hline
   9  & 0.8355 & 0.1746 & 0.9497 & 0.1916    \\    \hline
   15 & 0.9622 & 0.2710 & 1.0281 & 0.3274    \\    \hline
   20 & 1.2178 & 0.3997 & 1.3352 & 0.4887    \\    \hline
   25 & 1.3732 & 0.5041 & 1.4637 & 0.6623    \\    \hline
\end{tabular}
\end{center}
\end{table}
The variation of IWF with lorentz boost $y$ for different $D$ values are shown in Figure-1. The figure is self explanatory, obviously indicating that the zero recoil condition, $\xi(y)=1$ at $y=1$ is maintained all-through. \\
\begin{figure}[!htbp]
    \centering
    \subfigure[ $\xi(y)$ vs $y$ for D meson]
    {
        \includegraphics[width=3.2 in]{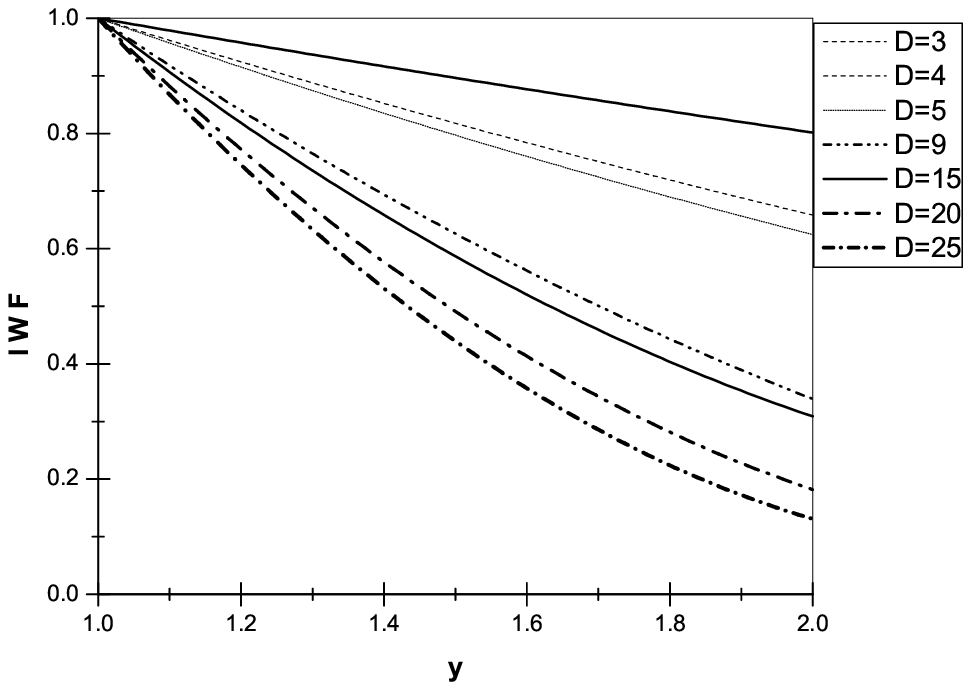}
        \label{fig:first_sub}
    }
        \subfigure[ $\xi(y)$ vs $y$ for B meson]
    {
        \includegraphics[width=3.2 in]{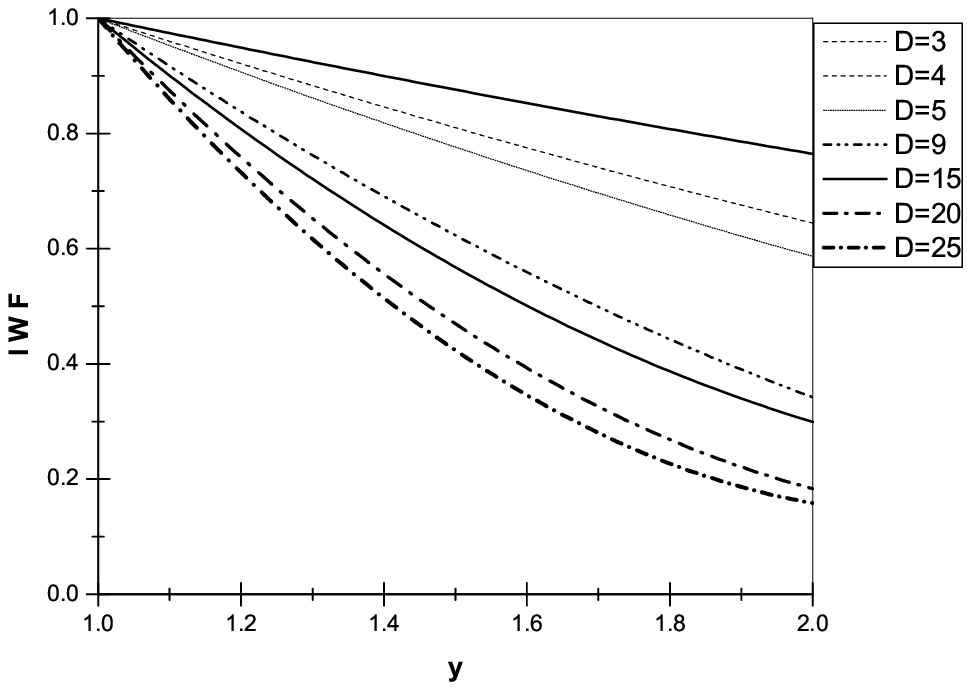}
        \label{fig:second_sub}
    }

    \caption{Variation of $\xi(y)$ with $y$ for different $D$.}
    \label{fig:sample_subfigures}
\end{figure}

For comparison of our results for $D=3$, the results of different standard models and collaborations in three dimensional QCD are shown in Table-4. \\
\begin{table}[!htbp]
\begin{center}
\caption{Results of slope and curvature of $\xi(y$) in different models and collaborations.}\label{cross}
\begin{tabular}{|c|r|r|}
  \hline
Model / collaboration &	Value of slope & Value of curvature \\
\hline \hline
 Ref [10] & 0.7936 & 0.0008 \\
 Le Youanc et al [36] & $\geq 0.75$ & $\geq 0.47$ \\
 Skryme Model [37] & 1.3 & 0.85 \\
 Neubert [38] & 0.82   & -- \\
 UK QCD Collab. [39]  & 0.83 & -- \\
 CLEO [40] & 1.67 & -- \\
 BELLE  [41] & 1.35 & -- \\
HFAG [42] & 1.17 $\pm 0.05$ & -- \\
Huang [43] & 1.35 $\pm 0.12$ & -- \\

  \hline
\end{tabular}
\end{center}
\end{table}

\section{Conclusion and remarks:}\rm

In this work, we have developed wave function for mesons in $d=D+1$ space-time dimension considering linear confinement term of Nambu-Goto potential as parent and L\"{u}scher term as perturbation. We have then further extended our analysis in finding the derivatives of Isgur-Wise function which are here obviously dimension dependent. \\
In calculating the slope and curvature of IWF, we have introduced cut-off value to the infinite upper limit of integrations to overcome divergences; this cut-off is found to be dimension dependent. We put forward our observation that our choice of specific cut-off value in these calculations does not compromise with the accuracy of our results. This will be more established, if we consider the asymptotic form of Airy's function (at $D=3$)[51]:
\begin{equation}
A_i[\varrho]_{asympt} \sim \frac{ \exp{(-\frac{2}{3}\varrho^{3/2}})}{2\sqrt{\pi}\varrho^{1/4}}
\end{equation}
With this asymptotic form (for $D=3$)we have also calculated the derivatives of $\xi(y)$ considering limit of integration from $r_0$ to $\infty$. The results for different cut-off values are shown in Table-5.
\begin{table}[!htbp]
\begin{center}
\caption{Values of  $\rho^2$ and C with asymptotic form of Airy's function.}\label{centre}
\begin{tabular}{|c|c|c|}
  \hline
  $r_0$ value & $\rho^2$ (asymptotic ) & C(asymptotic)\\
  \hline \hline
   5 & $4.6 \times  10^{-9}$ & $1.6 \times 10^{-9}$ \\
   \hline
   6 & $5.027 \times  10^{-11}$ &	$2.464 \times 10^{-11}$ \\
   \hline
   7& $3.56 \times  10^{-13}$ &	$2.345 \times 10^{-13}$ \\
   \hline
   8&  $1.695 \times  10^{-15}$ 	& $ 7.028 \times 10^{-15}$ \\
   \hline
   9&  $5.248 \times  10^{-18}$	& $6.597 \times 10^{-15}$ \\
   \hline
   10& $2.92 \times  10^{-19}$	& $2.78 \times  10^{-19}$ \\
    \hline
\end{tabular}
\end{center}
\end{table}
The values of $\rho^2$ and $C$ are exceptionally low here as compared to our calculated values thus justifying our consideration of restricting the upper limit of integration to some reasonable finite value. \\
While studying the compatibility of our results for $D=3$ with the standard results of different models in $3$ dimensional QCD (Table-4), we find that, our results are comparatively lower than the expectations. However,while comparing our results with that of our previous work [53], we put forward our comment that in ref[53] the results are exceptionally higher than the expectations and in our present approach we have overcome this limitation. Although our results are comparatively lower than the expectations, we find that $\rho^2$ and $C$ values gradually go on increasing with the increase of dimension $D$,  as is evident from Table-3, which is further confirmed from Figure-2. \\
\begin{figure}[!htbp]
    \centering
    \subfigure[ $\rho^2$ vs $D$ for D meson]
    {
        \includegraphics[width=3.2 in]{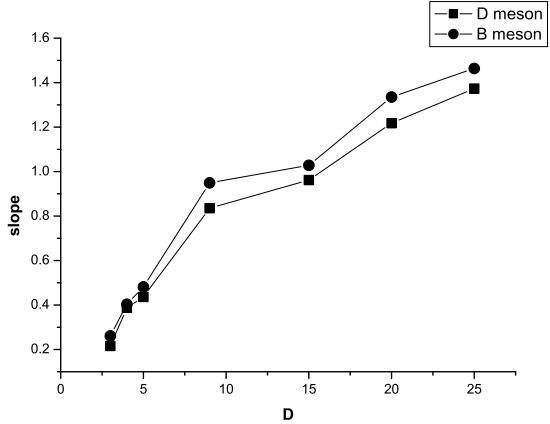}
        \label{fig:first_sub}
    }
        \subfigure[ $\rho$ vs $D$ for B meson]
    {
        \includegraphics[width=3.2 in]{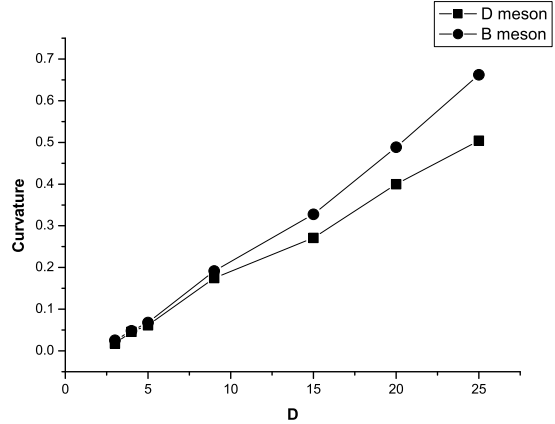}
        \label{fig:second_sub}
    }

    \caption{Variation of $\rho^2$ and $C$ with $D$ for D and B mesons.}
    \label{fig:sample_subfigures}
\end{figure}

Lastly, we make the following comments on our present formalism and calculations. \\
\begin{enumerate}
    \item In this work, we have developed meson wave function with an higher dimensional outlook and made studies of IWF and its derivatives within such approach.
   \item We opt for the quantum mechanical perturbation technique in deducing the meson wave function, due to the constraint in getting exact analytic solution of Schrodinger equation involving linear plus Coulombic potential.
  \item As far as numerical accuracy is concerned, the present perturbation approach appears short of the numerical solution of Schrodinger equation or calculations of Lattice QCD. Still, we believe that the potential model approach gives sufficiently reasonable physical insight into the problem.
  \item Lastly, terms $A_1(r,D),A_2(r,D),A_3(r,D)$ etc. of the second infinite series in the meson wave function are truncated up to $A_5(r,D)$ in our calculation. Improvement in the numerical analysis can be carried out by considering higher terms in both infinite polynomial series involved in the wave function.
  \end{enumerate}

\paragraph{Acknowledgement :\\ }
\begin{flushleft}
\emph{One of the authors (SR) acknowledges the support of University Grants Commission in terms of fellowship under Faculty Development Programme, to pursue research work at Gauhati University.}
\end{flushleft}

\appendix
\numberwithin{equation}{section}
\begin{center}
\section{Appendix}
\end{center}

\begin{center}
\textbf{\emph{Determination of equation (29) using Dalgarno's method of perturbation.}}
\end{center}

From equation (28), we have-
\begin{equation}
R_1(r)=\varrho_1 ^{m} r^{\frac{1-D}{2}+m} F(r) A_i[\varrho]
\end{equation}
We obtain-
\begin{eqnarray}
R_1 ^{\prime}(r)=\varrho_1 ^{m} r^{\frac{1-D+2m}{2}}[\frac{1-D+2m}{2} \frac{1}{r} F(r) Ai[\varrho] + F^{\prime}(r) Ai[\varrho] + \varrho_1 F(r)Ai^{\prime}[\varrho]]  \\
R_1 ^{\prime\prime}(r)=\varrho_1^{m} r^{\frac{1-D+2m}{2}}[\frac{1-D+2m}{2}(\frac{1-D+2m}{2}-1)\frac{1}{r^2}Ai[\varrho]F(r)+2\frac{1-D+2m}{2}\frac{1}{r}F^{\prime}(r)Ai[\varrho]+ \nonumber \\
 +2\frac{1-D+2m}{2}\frac{1}{r}\varrho_1 F(r)Ai^{\prime} [\varrho]+F^{\prime\prime}(r)Ai[\varrho]+2 \varrho_1 F^{\prime}(r)Ai^{\prime}[\varrho]+\varrho_1^{2}F(r)Ai^{\prime\prime}[\varrho]
\end{eqnarray}

We take :
\begin{eqnarray}
Ai^{\prime}[\varrho]= Z(r) Ai[\varrho] \;\;so\;\; that\;\; \\
Ai^{\prime\prime}[\varrho]=Z^{2}(r)Ai[\varrho]+Z^{\prime}(r)Ai[\varrho]
\end{eqnarray}
We then have-
\begin{eqnarray}
R_1^{\prime}(r)=\varrho_1^{m} r^{\frac{1-D+2m}{2}}Ai[\varrho][\frac{1-D+2m}{2} \frac{1}{r} F(r) + F^{\prime}(r)+ \varrho_1 F(r)Z]  \\
R_1^{\prime\prime}(r)=\varrho_1^{m} r^{\frac{1-D+2m}{2}}Ai[\varrho][\frac{1-D+2m}{2}(\frac{1-D+2m}{2}-1)\frac{1}{r^2}F(r) + 2\frac{1-D+2m}{2}\frac{1}{r}F^{\prime}(r)  \nonumber \\
 +2\frac{1-D+2m}{2}\frac{1}{r} \rho F(r) Z +F^{\prime\prime}+ 2 \varrho_1 F^{\prime}(r) Z+ \varrho_1^{2}F(r) Z^{2} + \varrho_1^{2} F(r) Z^{\prime}]
\end{eqnarray}
Putting equations (A.1), (A.6) and (A.7) in equation(27), we obtain:

\begin{flushleft}
\begin{eqnarray}
F^{\prime\prime}(r)+F^{\prime}(r) K_1(r,D) +F(r) K_2(r,D) -2\mu (\sigma r-E)F(r)= \nonumber \\
-2\mu (\frac{\gamma}{r}+W^{\prime})\;\;with \\
K_1(r,D)=2\frac{1-D+2m}{2}\frac{1}{r} +2\varrho_1 Z +\frac{D-1}{r}\\
K_2(r,D)= \frac{1-D+2m}{2}(\frac{1-D+2m}{2}-1)\frac{1}{r^2} +2\frac{1-D+2m}{2}\frac{1}{r}\varrho_1 Z  \nonumber \\
+ \varrho_1^2 (Z^2 +Z^{\prime})+ \frac{D-1}{r}\frac{1-D+2m}{2}\frac{1}{r}+ \frac{D-1}{r} \varrho_1 Z
\end{eqnarray}
\end{flushleft}

We take:
\begin{eqnarray}
Z(r)= \frac{C_1(r)}{r} \; \;and \\
Z^{2}(r) + Z^{\prime}(r)=\frac{C_2(r)}{r^2}
\end{eqnarray}
Our $K_1(r,D)$ and $K_2(r,D) $ transform to :
\begin{eqnarray}
K_1(r,D)=[2\frac{1-D+2m}{2}+(D-1)+2\varrho_1 C_1(r)]\frac{1}{r}= M_1(r,D) \frac{1}{r} \\
K_2(r,D)=[\frac{1-D+2m}{2}(\frac{1-D+2m}{2}-1) - (D-1)\frac{1-D+2m}{2} \nonumber \\
+ (2\frac{1-D+2m}{2} +(D-1))\varrho_1 C_1(r) +\varrho_1^{2}C_2(r)]\frac{1}{r^2}=M_2(r,D) \frac{1}{r^2}
\end{eqnarray}
where-
\begin{eqnarray}
M_1(r,D)=2\frac{1-D+2m}{2}+(D-1)+2 \varrho_1 C_1(r) \\
M_2(r,D)= \frac{1-D+2m}{2}(\frac{1-D+2m}{2}-1) - (D-1)\frac{1-D+2m}{2} \nonumber \\
+ (2\frac{1-D+2m}{2} +(D-1))\varrho_1 C_1(r) +\varrho_1^{2}C_2 (r)
\end{eqnarray}

It is to be noted that at $D=3$,
\begin{eqnarray}
K_1(r)=2\varrho_1C_1(r) \frac{1}{r} \\
K_2(r)=\varrho_1^{2}C_2(r) \frac{1}{r^2}
\end{eqnarray}

And equation (A.8) becomes:
\begin{equation}
F^{\prime\prime}(r) + 2\varrho_1 C_1(r) \frac{1}{r}F^{\prime}(r)+ \varrho_1^{2} C_2(r) \frac{1}{r^2}F(r) - 2\mu (\sigma r -E )F(r)= -2 \mu (\frac{\gamma}{r}+W^{\prime})
\end{equation}
which is similar to the equation obtained with 3-dimensional QCD potential.
We take:
\begin{equation}
F(r,D)=\sum_l A_l(r,D) r^{l}
\end{equation}
so that -
\begin{eqnarray}
F^{\prime}(r,D)= l\sum_l A_l(r,D) r^{l-1} \\
F^{\prime\prime}(r,D)=l(l-1)\sum_l A_l(r,D) r^{l-2}
\end{eqnarray}
Appling equations (A.20-A.22) in equation (A.8), we obtain:
\begin{equation}
\sum_l A_l(r,D) [l(l-1)+M_1(r,D) l +M_2(r,D) ] r^{l-2} +2\mu E \Sigma_l A_l(r,D) r^{l} -2\mu \sigma \sum_l A_l(r,D) r^{l+1} = -\frac{2\mu\gamma}{r} -2\mu W^{\prime}
\end{equation}
Equating power of $r^{-2}$ on both sides of equation (A.23), we get $ A_0 M_2 = 0 $, which imply that $ A_0 =0 $. \\
Further, equating powers of $ r^{-1}, r^{0}, r , r^{2}$  and  $r^{3}$ of equation (A.23), we obtain :
\begin{eqnarray}
A_1(r,D)= - \frac{2\mu \gamma}{M_1 +M_2} \\
A_2(r,D)= -\frac{2\mu W^{\prime}}{2+2M_1+M_2}\\
A_3(r,D)= \frac{4 \mu^{2} E \gamma}{(6+3M_1+M_2)(M_1+M_2)}\\
A_4(r,D)= \frac{4\mu^{2}E W^{\prime}(M_1+M_2)-4\mu^{2}\sigma\gamma (2+2M_1+M_2)}{(12+4M_1+M_2)(2+2M_1+M_2)(M_1+M_2)}\\
A_5(r,D)= - \frac{8\mu^{3} E^{2} \gamma(2+2M_1+M_2)+4\mu^{2}\gamma W^{\prime}(6+3M_1+M_2)(M_1+M_2)}{(20+5M_1+M_2)(6+3M_1+M_2)(2+2M_1+M_2)(M_1+M_2)}
\end{eqnarray}
With these expressions of  $A_1(r,D),A_2(r,D),A_3(r,D)$ etc we can now construct $F(r,D)$ as:
\begin{equation}
F(r,D)=A_1(r,D) r +A_2(r,D) r^{2} +A_3(r,D) r^{3} +A_4(r,D) r^{4}+A_5(r,D) r^{5} +\cdots\cdots\cdots
\end{equation}

And our radial wave function $R_1(r,D)$ comes out to be:
\begin{equation}
R_1(r,D)=r^{\frac{1-D}{2}} (\varrho_1 r)^{m} F(r,D)Ai[\varrho]
\end{equation}
The perturbed wave function $\Psi^{\prime}(r)$ and total wave function $\Psi^{tot} (r)$ are constructed as below:
\begin{eqnarray}
\Psi^{\prime}(r,D)=N_1r^{\frac{1-D}{2}} (\varrho_1 r)^{m}[A_1 (r,D) r +A_2(r,D) r^{2} +A_3(r,D) r^{3} +\cdots\cdots\cdots]Ai[\varrho_1 r-\varrho_0]\\
As,\;\;\Psi^{tot} (r,D)= \Psi^{0}(r,D)+ \Psi^{\prime}(r,D) \nonumber \\
= N_1r^{\frac{1-D}{2}} (\varrho_1 r)^{m}[1+ A_1(r,D) r +A_2 (r,D)r^{2} +A_3 (r,D)r^{3} +\cdots\cdots\cdots]Ai[\varrho_1 r-\varrho_0]
\end{eqnarray}

\numberwithin{equation}{section}
\begin{center}
\section{Appendix}
\end{center}

\begin{center}
\emph{\textbf{Calculation of $C_1(r)$ and $C_2(r)$ : }}
\end{center}

The Airy's infinite series as a function of $\varrho =\varrho_1 r - \varrho_0 $ can be expressed as [51] :
\begin{eqnarray}
Ai[\varrho_1 r - \varrho_0] = a_1[1+\frac{(\varrho_1 r - \varrho_0)^3}{6}+\frac{(\varrho_1 r - \varrho_0)^6}{180}+\frac{(\varrho_1 r - \varrho_0)^9}{12960}+...]- \nonumber \\
 b_1[(\varrho_1 r - \varrho_0) +\frac{(\varrho_1 r - \varrho_0)^4}{12}+\frac{(\varrho_1 r - \varrho_0)^7}{504}+\frac{(\varrho_1 r - \varrho_0)10}{45360}+...]
\end{eqnarray}

\begin{flushright}
with $ a_1=\frac{1}{3^{2/3}\Gamma(2/3)}=0.3550281$ and $b_1=\frac{1}{3^{1/3}\Gamma(1/3)}=0.2588194$.\\
\end{flushright}
To find $ C_1(r)$ and $C_2(r) $ , we take truncated Airy series up to lowest order, so that, we have :
\begin{eqnarray}
Z(r)=\frac{C_1(r)}{r} = \frac{Ai^{\prime}(\varrho_1 r-\varrho_0)}{Ai(\varrho_1 r-\varrho_0)} \\
=\frac{-b_1 \varrho_1}{a_1-b_1(\varrho_1 r-\varrho_0)} \\
=\frac{b_1 \varrho_1}{b_1(\varrho_1 r-\varrho_0)-a_1} \\
=\frac{1}{r}[1-\frac{a_1+b_1\varrho_0}{b_1\varrho_1}\frac{1}{r}]^{-1} \\
=\frac{1}{r}[1-\frac{k}{r}]^{-1}
=\frac{1}{r}(1+\frac{k}{r}) \\
Therefore,\; C_1(r)=1+\frac{k}{r} \\
with\; k=\frac{a_1+ b_1 \varrho_0}{b_1 \varrho_1}
\end{eqnarray}
Also,
\begin{eqnarray}
Z^{2}(r)=\frac{1}{r^{2}}(1+\frac{2k}{r}+\frac{k^{2}}{r^{2}})=\frac{1}{r^{2}}+\frac{2k}{r^{3}}+\frac{k^{2}}{r^{4}} \\
and\; \; Z^{\prime}(r)=-\frac{1}{r^{2}}-\frac{2k}{r^{3}} \;\;so\;that,\\
\frac{C_2(r)}{r^{2}}=Z^{2}(r)+Z^{\prime}(r)=\frac{k^{2}}{r^{4}}
\end{eqnarray}
We thus obtain-
\begin{equation}
 C_2(r)=\frac{k^{2}}{r^{2}}
\end{equation}

\begin{center}
*****
\end{center}


\begin{thebibliography}{99}

\bibitem{1} 	J. D. Bjorken, SLAC-PUB-5278 (1990).
\bibitem{2}	    A. Martin, Phy Lett. B 93, 338(1980).
\bibitem{3}  	E. Eichten, K. Gottfried, T. Kinoshita, K. D. Lane, and T.-M. Yan, ``Charmonium: Comparison with experiment," Phys. Rev. D21 , 203(1980).
\bibitem{4} 	J.L.Richardson, Phy. Lett. B 82,272(1979).
\bibitem{5} 	C. Quigg and J. L. Rosuer, Phy. Lett. B 71,153(1977).
\bibitem{6}	    A. K. Gathak and S. Lokanathan in ``Quantum Mechanics"; McGraw Hill(1997), pp-291.
\bibitem{7} 	D. K. Choudhury, P. Das, D. D. Goswami and J. N. Sarma; Pramana- J of Phys. 44, 519 (1995).
\bibitem{8} 	D. K. Choudhury and N. S. Bordoloi, MPLA, 24, 443-451(2009).
\bibitem{9}	    D. K. Choudhury and N S Bordoloi; Int. J. Mod. Phys A 15, 3667(2000).
\bibitem{10}	B. J. Hazarika , K. K. Pathak and D. K. Choudhury, MPLA Vol 26, No 21,1547-1554 (2011).
\bibitem{11}  	S.Roy  and D.K.Choudhury, MPLA Vol \textbf{27} No 20,1250110 (2012).
\bibitem{12}  	M. Luscher, Nucl. Phy. B , 180,317 (1981)
\bibitem{13}  	J. F. Arvis, Phy. Lett B, 127 , 106(1983).
\bibitem{14}  	A. Antillon et al, Phy. Rev. D , Vol 49, No. 4 ( 1994).
\bibitem{15}	Y. Nambu: Phys. Lett. B 80, 372 (1979).
\bibitem{16}  	M. Luscher, K. Symanzik and P.Weisz: Nucl. Phys. B 173, 365 (1980).
\bibitem{17}    A. Yamamato, H. Suganuma and H. Iida, Phys.Lett.B \textbf{664},129 (2008).
\bibitem{18}    Cardoso N. and Bicudo P. , Phy. Rev. D \textbf{85}(2012)077501 (arXiv:1111.1317 [hep-latt]).
\bibitem{19}	A. Chaterjee, Phys. Rev A 35, 2722(1987).
\bibitem{20}	S. H. Dong et al, Found. Phys. Lett, 12, 465 (1999).
\bibitem{21}	S. H. Dong, Found. Phys. Lett 15,385(2002).
\bibitem{22}	S. H. Dong, Physica Scripta, Vol 64,273-276(2001).
\bibitem{23}	Chen Gang, Chinese Physics, Vol 14 No. 6, 1075-1076 ( 2005 ).
\bibitem{24}	H. Hassanabahi \emph{et al}, Commun. in Th. Phys. \textbf{55},541(2011).
\bibitem{25} 	H.Hassanabahi \emph{et al}, Chin. Phy. Lett. Vol \textbf{29}, No.2,020303(2012).
\bibitem{26}	 H.Hassanabahi \emph{et al}, Int. J. of Quant. Chem. DOI:10.1002/qua.24064 (2012).
\bibitem{27} 	H.Hassanabahi , Commun. in Th. Phys. \textbf{55},303(2011).
\bibitem{28} 	H.Hassanabahi\emph{et al}, Eur. Phy. J. Plus Vol \textbf{127}, NO. 5,51(2012).
\bibitem{29} 	H.Hassanabahi \emph{et al}, MPLA 24,1043 (2009).
\bibitem{30}	P. J. Olver and C. Shakiban in ``Applied Mathematics" (2004),pp1039-1042.
\bibitem{31}	A. Dalgarna and J. T. Lewis, Proc. R. Soc. A, 233,70(1955).
\bibitem{32} 	T. K. Nandi et al, J Phy A: Math Gen 29, 1101 (1996).
\bibitem{33} 	N. Isgur and M B Wise, Phys Lett. B 232 , 113 (1989).
\bibitem{34} 	N. Isgur and M B Wise,Phy. Rev. Lett 66, 1130 (1991).
\bibitem{35} 	Kacper Zalewski, CERN-TH,6891 (1993).
\bibitem{36}	A. Le Yaouanc, L Oliver et al; Phys Lett. B 365,319(1996).
\bibitem{37}	E. Jenkins, A. Manahar, M. B. Wise; Nucl. Phys B , 396; 38(1996).
\bibitem{38}	M. Neubert, Phys Lett B 264; 455(1991).
\bibitem{39}	UKQCD Collab. , K C Bowler et al; Nucl Phys B, 637, 293(2002).
\bibitem{40}	CLEO Collab., J Bartel et al, Phys Rev Lett 82, 3746(1999).
\bibitem{41}	BELLE collab, K Abe et al, Phys Lett B, 526, 258(2002).
\bibitem{42}	Heavy Flavor Averaging Group (HFAG), hep-ex/08081297(2009).
\bibitem{43}	Ming-Qiu Huang et al, Phys.Letts. B 629,27-32(2005).
\bibitem{44}	H. Hassanabadi et al, Int. J. of the Phys. Sc. Vol 6(3),583-586 (2011) .
\bibitem{45}	Abramowitz and Stegun ; ``Handbook of Mathematical Functions" , 10th ed ,National Bureau of Standards, US (1964)pp446.
\bibitem{46} 	S. Nouri , Phy Rev A 60(2),1702(1999).
\bibitem{47} 	H. Politzer and M. Wise, Phys. Lett. B 206, 681(1988).
\bibitem{48}	H. Georgi, Phys. Lett., B240,447(1990).
\bibitem{49}	F. E. Close and A Wanbach, Nucl. Phy B 412, Issue 1-2( 1994) p 169-180.
\bibitem{50}	Yuan-Ben Dai, C S Huang and H Y Jim, Z Phys C 56, 707(1992).
\bibitem{51}	Valle´e O and Soares M,``Airy Functions and Applications in Physics",London: Imperial College Press( 2004 )pp7.
\bibitem{52}    S. Roy, N S Bordoloi and D. K. Choudhury: arXiv:1209.6121[hep-ph].
\bibitem{53}    S Roy, B J Hazarika and D K Choudhury, Phys. Scr 86 , 045101(2012).
\bibitem{54}    Noura Eduarda, David G. Henderson in ``Experiencing Geometry: On plane and sphere", Prentice Hall (1996).
\bibitem{55}    H. Hassanabadi et. al., Commun. Theor. Phys .,Vol. 56, No. 3, 423(2011).
\bibitem{56}    H. Hassanabadi et. al., Annalen der Physik, vol. 523, No. 7,566(2011).
\bibitem{57}    M. Neubert, Phys. Rept. 245, 259(1994).
\bibitem{58}    M.G. Olsson and S. Veseli, arXiv:hep-ph/9702212.



\end{thebibliography}
\end{document}